\documentclass[10pt, twocolumn, pre, aps, superscriptaddress, showpacs]{revtex4-1}
\usepackage{amsmath, graphicx, subfigure, tikz}
\begin{document}

\title{Metastable Reverse-Phase Droplets within Ordered Phases:\\
Renormalization-Group Calculation of Field and Temperature\\
Dependence of Limiting Size}
    \author{Ege Eren}
    \affiliation{Department of Electrical Engineering, Bo\u{g}azi\c{c}i University, Bebek, Istanbul 34342, Turkey}
    \author{A. Nihat Berker}
    \affiliation{Faculty of Engineering and Natural Sciences, Kadir Has University, Cibali, Istanbul 34083, Turkey}
    \affiliation{Department of Physics, Massachusetts Institute of Technology, Cambridge, Massachusetts 02139, USA}
%    \pacs{75.10.Nr, 05.10.Cc, 64.60.De, 75.50.Lk}

%05.10.Cc    Renormalization group methods
%64.60.ae    Renormalization-group theory
%64.60.De    Statistical mechanics of model systems
%75.10.Nr    Spin-glass and other random models
%75.50.Lk    Spin glasses and other random magnets

%64.60.Cn    Order-disorder transformations
%05.50.+q    Lattice theory and statistics (Ising, Potts, etc.)
%61.43.-j    Disordered solids
%75.10.Hk    Classical spin models

\begin{abstract}

Metastable reverse-phase droplets are calculated by
renormalization-group theory by evaluating the magnetization of a
droplet under magnetic field, matching the boundary condition with
the reverse phase and noting whether the reverse-phase magnetization
sustains.  The maximal metastable droplet size and the discontinuity
across the droplet boundary are thus calculated as a function of
temperature and magnetic field for the Ising model in three
dimensions. The method also yields hysteresis loops for finite
systems, as function of temperature and system size.

\end{abstract}
\maketitle

\section{Introduction: Calculated Sustainability of Metastable Droplets}

After outstanding success in the calculation of critical exponents
and in understanding the mechanism underlying the universality of
critical exponents, renormalization-group theory has been equally
successfully applied to global non-universal properties at and away
from critical points, such as entire thermodynamic functions,
discontinuities at first-order phase transitions, and entire
multicritical phase diagrams, e.g., leading all the way to global
spin-glass phase diagrams in the variables of temperature, bond
concentration, spatial dimensionality $d$, and the continuous
variation of chaos and its Lyapunov exponent inside spin-glass
phases \cite{Ilker2,Atalay}.  Such wide application has not yet been
reached in non-equilibrium systems.  On the other hand,
renormalization-group calculations have been applied to finite-sized
systems \cite{BerkerOstlund}.  As explained below, these
calculations can be used to obtain the properties of non-equilibrium
metastable droplets inside ordered phases and the hysteresis loops
of systems partitioned into domains.  In addition to the metastable
applications mentioned above, our calculations are applicable to
paint cratering due to surface defects \cite{Deutch}.

Non-equilibrium studies away from the critical point have been on
the droplet formation of the equilibrium phase, inside the
non-equilibrium metastable phase (after a deep quench).  For the
much studied Classical Nucleation Theory, see Refs.
\cite{BeckerDoring,Langer,Ryu,Stauffer,Gunton,Klein1,Klein2}. We
tackle here the converse problem, namely the existence of metastable
droplets of the non-equilibrium phase, inside the equilibrium phase.
We find that such droplets exist, and are clearly delimited by size,
magnetic field, and temperature, the latter two being a measure of
the thermodynamic distance to the first-order phase boundary where
the non-equilibrium phase becomes a coexisting phase.

\begin{figure}[ht!]
\centering
\includegraphics[scale=0.28]{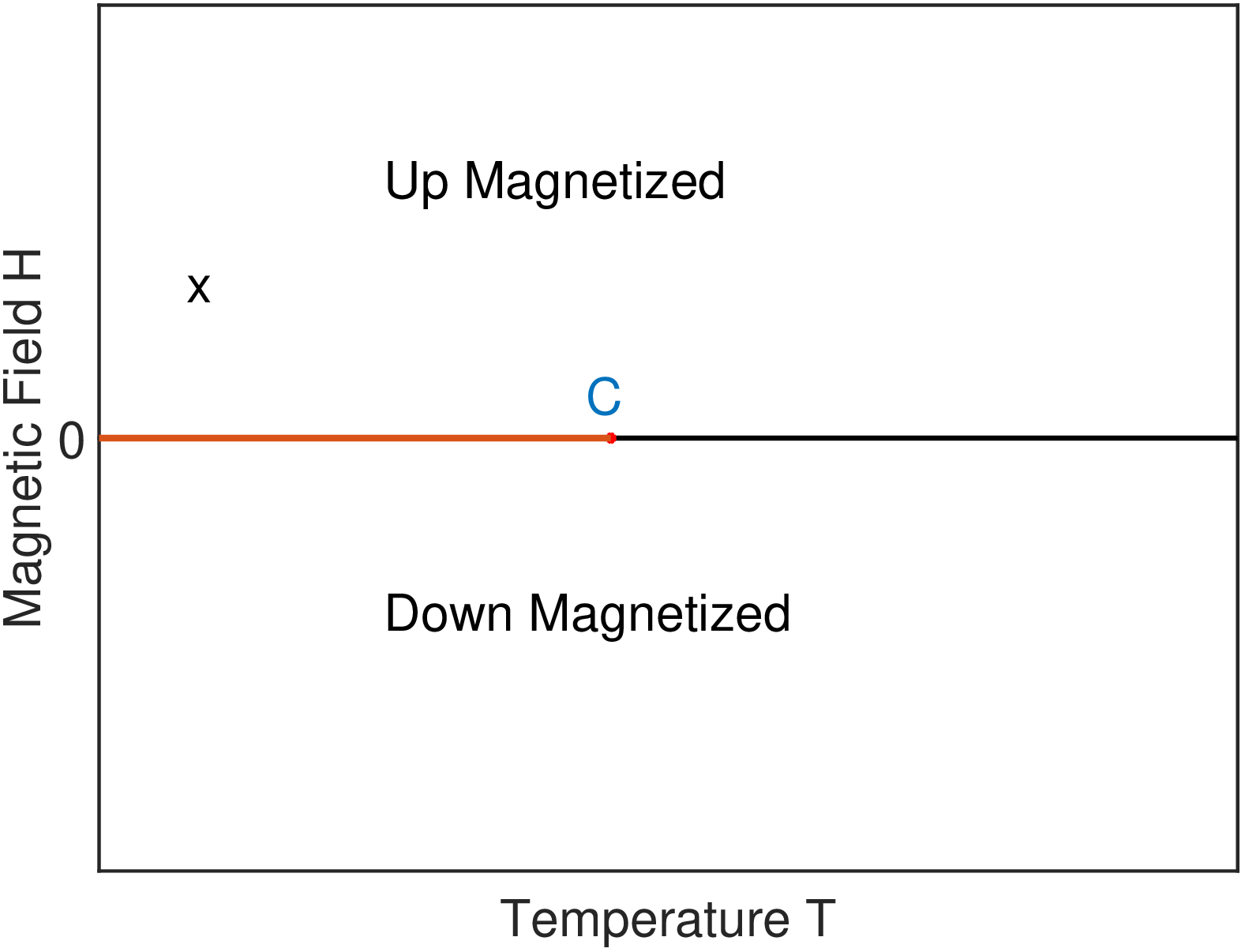}
\caption{Phase diagram of the ferromagnetic Ising model for $d > 1$.
The equilibrium phases are indicated.  For $T < T_C$ on the $H = 0$
line, there is a first-order phase transition between the
down-magnetized phase and the up-magnetized phase.  At the position
in the phase diagram marked by $\times$, the equilibrium phase is
the up-magnetized phase, but metastable droplets of the
down-magnetized phase can exist, depending on the
thermodynamic-variable distance from the phase boundary.}
\end{figure}

\begin{figure}[ht!]
\centering
\includegraphics[scale=0.3]{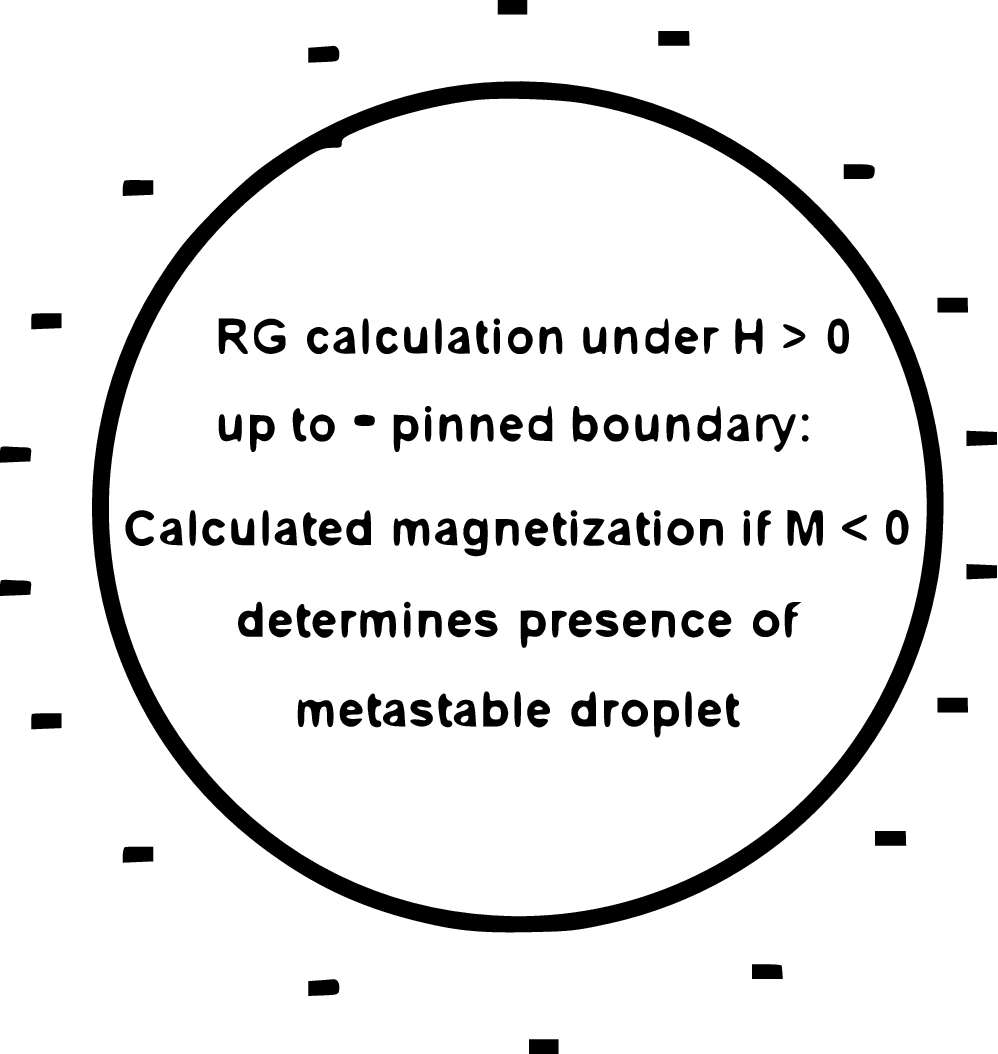}
\caption{The magnetization is calculated inside a droplet of size
$L$ at position $\times$ on Fig. 1, that is under the system-wide
applicable value of $H>0$, matching with the boundary condition of
down-pinned, $s_i = -1$ spins.  This boundary represents the
outermost layer of the would-be droplet.  If this calculation gives
$M < 0$ inside the droplet, the droplet exists. Note that our
calculation is for a cubic shape, under the Migdal-Kadanof
approximation, not the spherical object shown in this figure that
pedagogically explains our method.}
\end{figure}

\begin{figure}[ht!]
\centering
\includegraphics[scale=0.4]{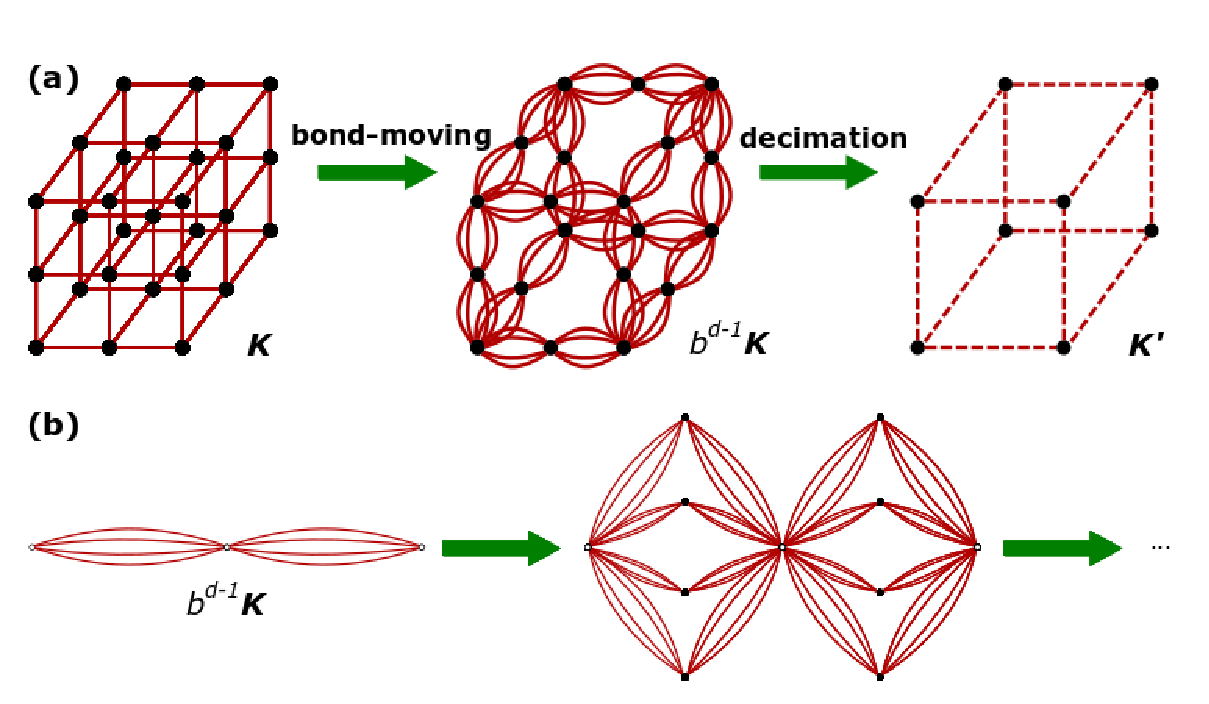}
\caption{(a) Migdal-Kadanoff approximate renormalization-group
transformation for the $d=3$ cubic lattice with the length-rescaling
factor of $b=2$. (b) Construction of the $d=3, b=2$ hierarchical
lattice for which the Migdal-Kadanoff recursion relations are exact.
The renormalization-group solution of a hierarchical lattice
proceeds in the opposite direction of its construction.}
\end{figure}

\section{Boundary-Conditioned Finite-System Calculation for the Presence of a Metastable Droplet}

Finite-system renormalization-group calculation \cite{BerkerOstlund}
can readily be adapted to metastable droplet viability.  Such a
droplet is a finite region of the opposite thermodynamic phase
persisting inside the equilibrium phase.  Even as a metastable
region, a droplet can exist up to a certain (critical) size,
depending on how far away, in the thermodynamic external (applied)
variables, the system is from the boundary in thermodynamic space
where the phase of the droplet becomes stable. The further away, the
smaller the maximal droplet size, up to a certain limit, beyond
which metastable droplets do not exist. This critical droplet size
and this limit of metastable droplet existence are obtained, based
on a microscopic statistical mechanical (renormalization-group)
calculation, in our current work.

We illustrate with the Ising model, defined by the Hamiltonian
\begin{equation}
-\beta \mathcal{H}= J \sum_{\langle ij \rangle} s_i s_j + H \sum_i
s_i\,,
\end{equation}
where $\beta=1/kT$, at each site $i$ of the lattice the spin $s_i =
\pm 1$, and $\langle ij \rangle$ denotes summation over all
nearest-neighbor site pairs. The bond is ferromagnetic, $J>0$, and
$J^{-1}$ is thus the temperature variable.  In three spatial
dimensions $(d=3)$, this system has the phase diagram shown in Fig.
1.  The equilibrium phases are indicated.  As is well known, for $T
< T_C$ on the $H = 0$ line, there is a first-order phase transition
between the down-magnetized phase with $ M = <s_i> < 0$ and the
up-magnetized phase with $M > 0$.  At the position in the phase
diagram marked by $\times$, where $H>0$, the equilibrium phase is
the up-magnetized phase with $M>0$, but metastable droplets of the
down-magnetized phase with $M<0$ can exist, depending on the value
of $H$, which gives the thermodynamic-variable distance from the
phase boundary.

\begin{figure}[ht!]
\centering
\includegraphics[scale=0.3]{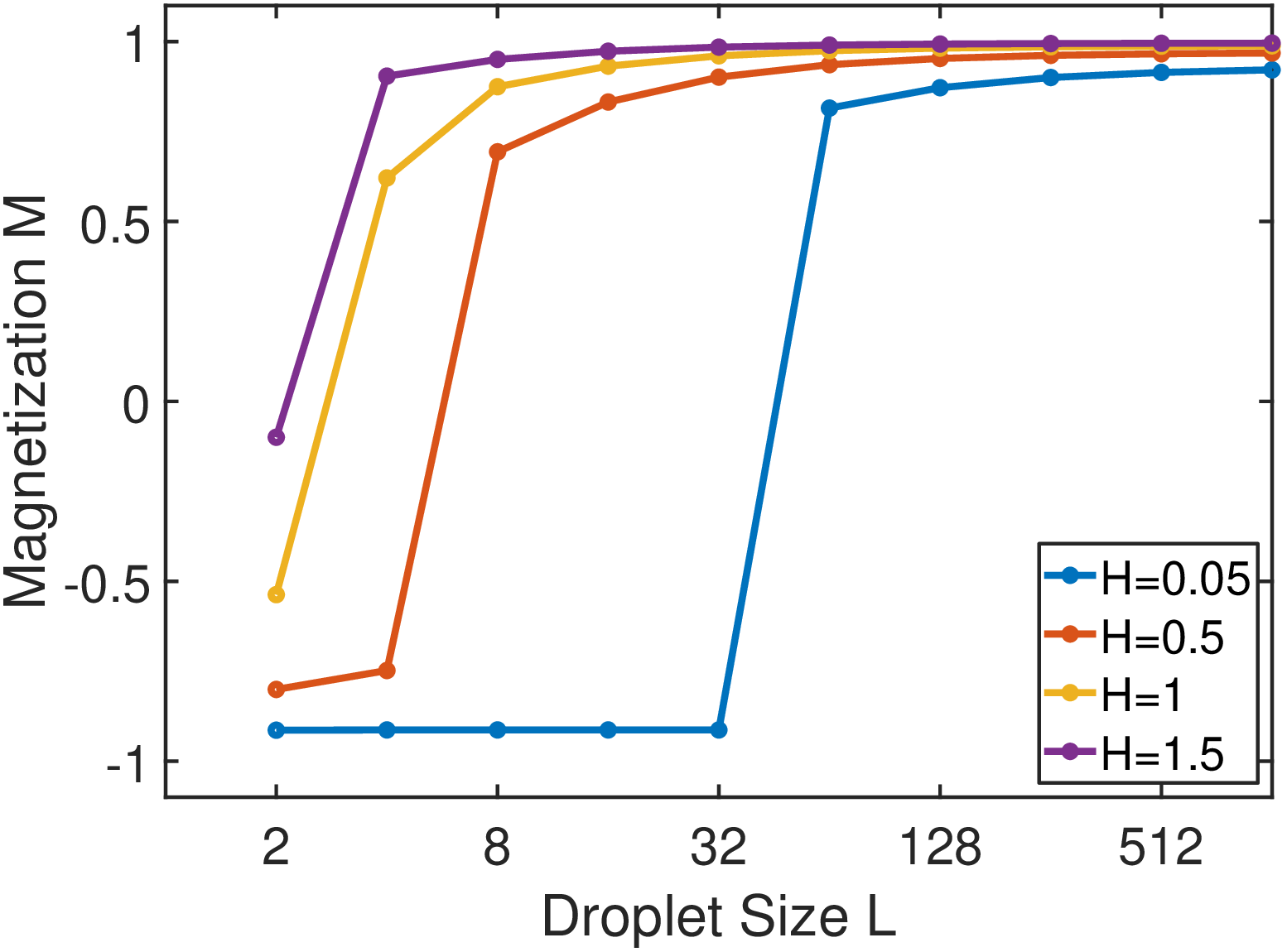}
\caption{Magnetization inside a would-be droplet of negative
magnetization immersed in the positive magnetization phase, as a
function of droplet size.  Results for different magnetic fields $H$
are shown, from right to left for $H =
0.05,.5,1,1.5,$ at temperature $J^{-1} = 5$. When the magnetization
calculated inside the droplet is positive, the droplet does not
occur.}
\end{figure}

\begin{figure}[ht!]
\centering
\includegraphics[scale=0.27]{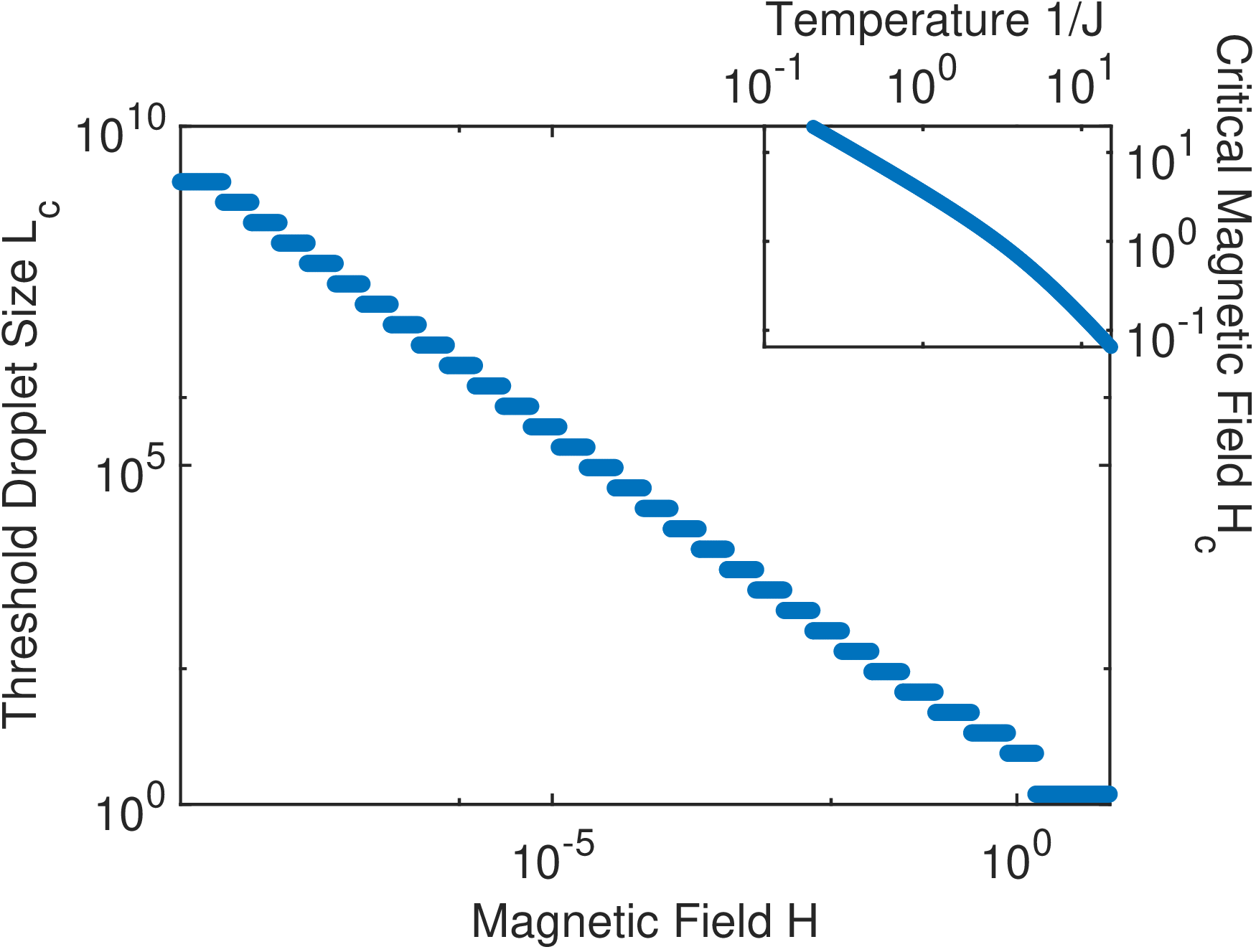}
\caption{Threshold sizes $L_C$ as a function of magnetic field $H$,
at temperature $J^{-1} = 5$.  The threshold droplet size $L_C$ goes
to infinity as $L_C \sim H^{-0.94}$ as the system approaches the
phase transition at $H=0$. Conversely, $L_C$ goes to zero as the
systems moves away from the phase transition.  Metastable droplets
do not occur for $H > H_{max}(J)$, which is seen to be $H_{max} =
1.60$ for temperature $T = J^{-1} = 5$. The inset gives $H_{max}$ as
a function of temperature.  $H_{max}$ diverges as $H_{max} \sim
T^{-1.05}$ as zero temperature is approached.  At a higher
temperature, there is a crossover to $H_{max} \sim T^{-1.84}$, as
seen on the right side of $T \simeq 1$ in the
inset.}
\end{figure}

The basis of our method is as follows:  Renormalization-group theory
enables us to calculate the thermodynamic properties, including
magnetization $M$ of finite and infinite systems.  Thus, we
calculate the magnetization inside a droplet of size $L$ at position
$\times$ on Fig. 1, that is under the system-wide applicable value
of $H>0$.  We match with the boundary condition of down-pinned, $s_i
= -1$ spins.  This boundary represents the outermost layer of the
would-be droplet.  If this calculation gives $M < 0$ inside the
droplet, the droplet exists.  If, in this calculation, the applied
field $H > 0$ overcomes the down-pinned boundary condition $s_i =
-1$ and $M > 0$, then the droplet of size $L$ does not exist.  Since
under renormalization-group transformation, the value of the
magnetic field increases, there is clearly a maximal value $L_C(H)$,
where $H$ refers to the initial (unrenormalized) magnetic fields, to
the metastable droplet size.

\section{Renormalization-Group Calculation}
For our calculation, we use the Migdal-Kadanoff approximation,
which, as shown in Fig. 3(a), consists in bond-moving followed by
decimation \cite{Migdal,Kadanoff}.  This much used approximation
gives, for the first renormalization-group transformation, the
recursion relations
\begin{equation}
K'=K'(J,H),\, h'=h'(J,H),\, G' = G'(J,H),
\end{equation}
where the primes refer to the renormalized quantities and $G$ is the
additive constant in the Hamiltonian automatically generated by the
renormalization-group transformation:
\begin{equation}
-\beta \mathcal{H}= \sum_{\langle ij \rangle} [K s_i s_j + h (s_i +
s_j) + G].
\end{equation}
In the latter equation, the magnetic field term is expressed in the
form that it takes after the first renormalization-group
transformation. After the first renormalization-group
transformation, the form of the Hamiltonian in Eq. (3) is conserved
and the recursion relations have the form
\begin{equation}
K'=K'(K,h),\, h'=h'(K,h),\, G'=b^d G + g(K,h).
\end{equation}

The recursion relations thus obtained from the Migdal-Kadanoff
approximation are exactly applicable to the exact solution of the
hierarchical lattice shown in Fig. 3(b)
\cite{BerkerOstlund,Kaufman1,Kaufman2}. Thus, a "physically
realizable", therefore robust approximation is used.  Physically
realizable approximations have been used in polymers
\cite{Flory,Kaufman}, disordered alloys \cite{Lloyd}, and turbulence
\cite{Kraichnan}.  Recent works using exactly soluble hierarchical
lattices are in Refs.
\cite{Monthius,Sariyer,Ruiz,Rocha-Neto,Ma,Boettcher5}.

Sites at different levels of a hierarchical lattice have different
coordination numbers.  For example, on the right side of Fig. 3(b),
two different levels are seen for the inner sites, with coordination
numbers $q = 8$ and $16$.  In our calculation, the magnetic field in
Eq. (1) is applied to sites in the lowest level of the hierarchy,
which are the least coordinated $(q = 8)$ and most numerous,
comprising $1-b^{-d} = 7/8$ of all the sites.

In the first renormalization-group transformation, going from Eq.
(1) to Eq. (3) with a scale change, the thermodynamic densities are
related by
\begin{equation}
\textbf{M} = b^{-d} \, \textbf{m} \cdot \textbf{U},
\end{equation}
where the densities $\textbf{M} \equiv [1, <s_is_j>, <s_i>]$ and
$\textbf{m} \equiv [1, <s_is_j>, <(s_i+s_j)>]$ respectively refer to
Eqs. (1) and (3). The matrix is $\textbf{U} =
\partial \textbf{k} /
\partial\textbf{K}$ with
$\textbf{k} \equiv [1, K, h]$ and $\textbf{K} \equiv [1, J,
b^{-d}H]$.  The factor $b^{-d}$ in the last vector accounts for the
fact that there are as many unrenormalized fields $H$ (applied only
to the lowest level of the hierarchy) as renormalized bonds $K$.

In the subsequent renormalization-group transformations, the
thermodynamic densities obey the recursion relation
\begin{equation}
\textbf{m} = b^{-d} \textbf{m'} \cdot \textbf{T},
\end{equation}
where the densities $\textbf{m} \equiv [1, <s_is_j>, <(s_i+s_j)>]$
are conjugate to the fields $\textbf{k} \equiv [1, K, h]$ and the
recursion matrix is $\textbf{T} = \partial \textbf{k'} /
\partial\textbf{k}$.  The densities are calculated by iterating
Eq.[6] until the boundary is reached and applying the boundary
condition in the right side of Eq. [6].

The calculation yields the phase diagram in Fig. 1, with the
critical temperature $J^{-1} = 15.3$, which sets the temperature
scale.

\begin{figure}[ht!]
\centering
\includegraphics[scale=0.28]{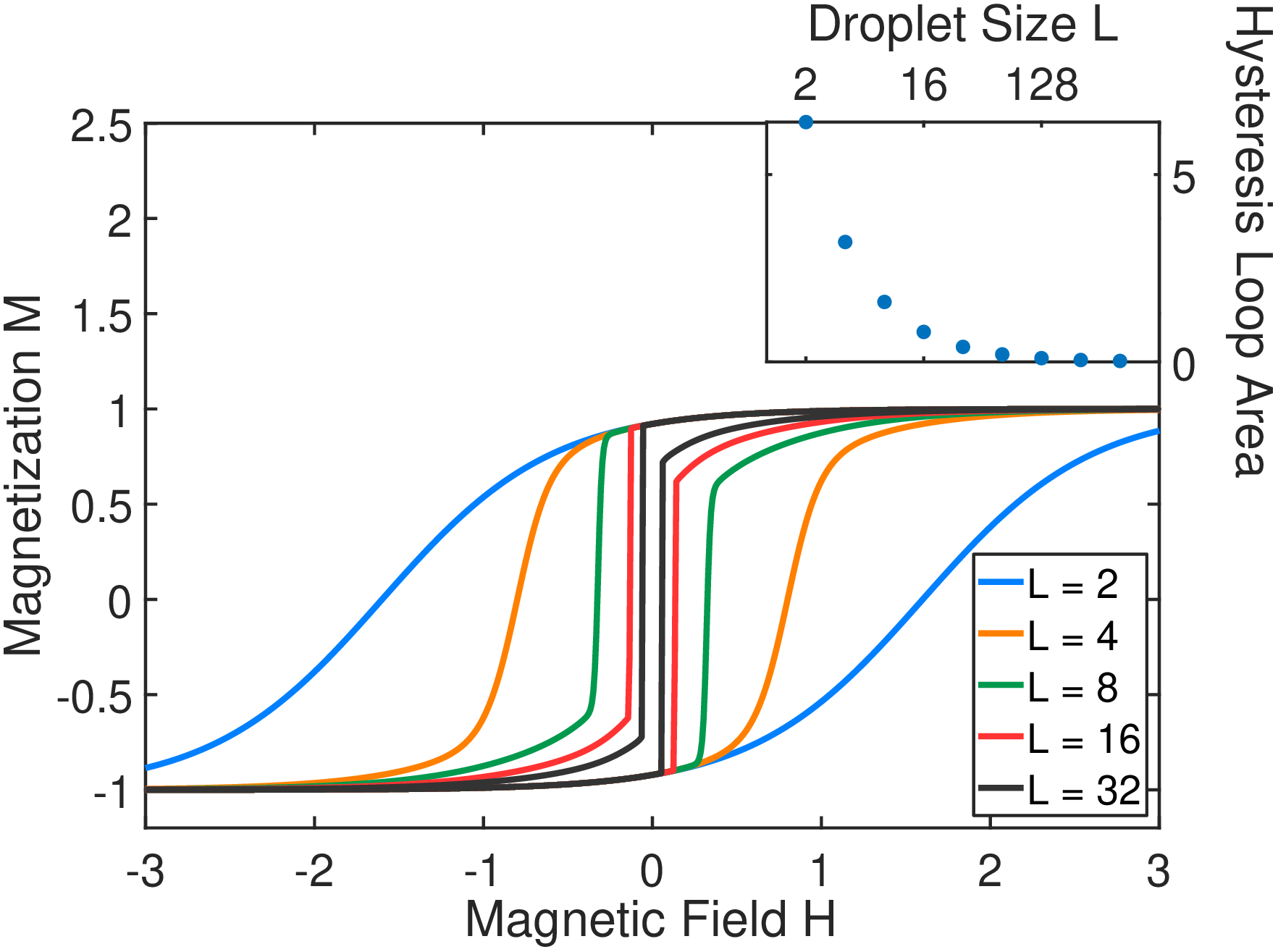}
\caption{Hysteresis loops at temperature $J^{-1} = 5$, for
differently sized systems $L = 2,4,8,16,32$ from the
sides to the middle. The loops get narrower as the system size
grows, leading to a single discontinuity of the first-order phase
transition of the infinite (thermodynamic limit) system. The inset
shows the hysteresis loop area as a function of system size. A
maximum loop area is reached when extrapolated to vanishing system
size.}
\end{figure}

\begin{figure}[ht!]
\centering
\includegraphics[scale=0.25]{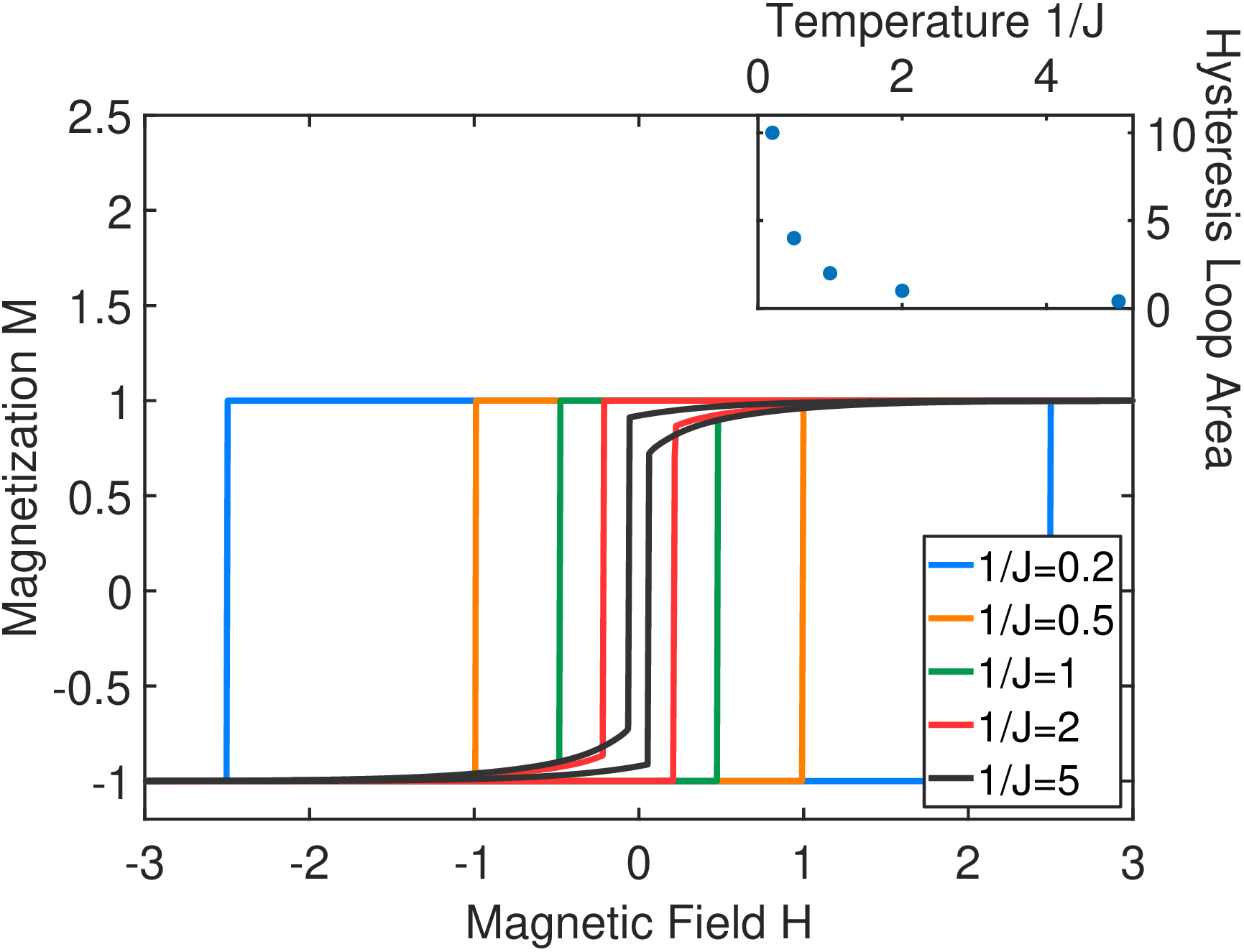}
\caption{Hysteresis loops at system size $L = 32$,
for different temperatures $J^{-1} = 0.2,0.5,1,2,5$ from the sides
to the middle. The loops get narrower as temperature increases. The
inset shows the loop area as a function of temperature.  The loop
area diverges as zero temperature is approached.}
\end{figure}

\section{Results: Calculated Metastable Droplet Size Thresholds}

Magnetizations, calculated as described above, inside a would-be
droplet of negative magnetization immersed in the positive
magnetization phase, as a function of size, are shown in Fig. 4.
Results for different magnetic fields $H$ are shown, at temperature
$J^{-1} = 5$.  When the magnetization calculated inside the droplet
is negative, the droplet-boundary match is self-consistent and the
droplet does occur.  It is seen in Fig. 4 that this droplet
occurrence terminates, with a discontinuity in the calculated
magnetization, at a threshold size $L_C$.  Here $L = b^n$ is the
linear dimension of the droplet, where $n$ is the number of
renormalization-group transformations needed for the many-times
renormalized spin nearest-separation to span the droplet.

Threshold sizes $L_C$, as a function of magnetic field $H$, are
shown in Fig. 5, at temperature $J^{-1} = 5$.  The threshold droplet
size $L_C$ goes to infinity as $L_C \sim H^{-0.94}$ as the system
approaches the phase transition at $H=0$. Conversely, $L_C$ goes to
zero as the systems moves away from the phase transition. Metastable
droplets do not occur for $H > H_{max}(J)$, which is seen to be
$H_{max} = 1.60$ for temperature $T = J^{-1} = 5$. The inset gives
$H_{max}$ as a function of temperature.  $H_{max}$ diverges as
$H_{max} \sim T^{-1.05}$ as zero temperature is approached.  At a
higher temperature, there is a crossover to $H_{max} \sim
T^{-1.84}$, as seen on the right side of $T \simeq
1$ in the inset. In Fig. 5, the discreteness of the droplet size
values is due to the discrete nature of the (multiplicative) length
rescaling factor $b=2$ of the renormalization-group transformation.

\section{Results: Calculated Hysteresis Loops}

Our current calculation can also readily be adapted to a hysteresis
loop calculation.  Hysteresis loops occurs in systems composed of
finite microdomains.  As the magnetization measurements/calculations
are made, the system carries a memory of the previous
measurement/calculation.  In the present case, this is accomplished
by keeping down-pinned boundary spins as the system is scanned in
the increasing $H$ direction and conversely up-pinned boundary spins
as the system is scanned in the decreasing $H$ direction.

Thus, hysteresis loops at temperature $J^{-1} = 5$, for differently
sized systems, are shown in Fig. 6.  The loops get narrower as the
system size grows, leading to a single discontinuity of the
first-order phase transition of the infinite (thermodynamic limit)
system. The inset shows the hysteresis loop area as a function of
system size. A maximum loop area is reached when extrapolated to
vanishing system size.

Hysteresis loops at system size $L = 32$, for
different temperatures, are shown in Fig. 7.  The loops get narrower
as temperature increases, as the spin-spin correlations and
specifically the correlations to the "memory" spins (in this case
the boundary spins) decreases.  The inset shows the loop area as a
function of temperature.  The loop area diverges as zero temperature
is approached.

\section{Conclusion}

We have calculated metastable droplets inside a stable thermodynamic
phase, using renormalization-group theory.  We find the droplet
existence is clearly delimited by droplet size and thermodynamic
distance to the stability of the phase inside the droplet.  The
method was also adapted to hysteresis loop calculation, yielding
loop areas as a function of domain size and temperature.

We note that the inside bulk of the metastable droplet in our study
increases the free energy with respect to the thermodynamically
stable phase, as does the boundary between the metastable droplet
and the stable phase.  The phenomenological explanation for the
occurrence of the metastable droplet is that there exists a free
energy barrier to reverting this droplet.  With increasing magnetic
field, the free energy of the bulk of the droplet increases,
reaching, together with the boundary free energy, the level of the
free energy barrier, thus turning the barrier into a shoulder and
causing the droplet to revert to the stable phase. Finally, a Monte
Carlo calculation to check our results would be an interesting
called-for study.

\begin{acknowledgments}
We are grateful to E.C. Artun for useful comments. A.N.B. thanks Prof. David Turnbull for encouragement of this work,
some time ago. Support by the Academy of Sciences of Turkey (T\"UBA)
is gratefully acknowledged.
\end{acknowledgments}

\end{document}